# Assembly of "3D" plasmonic metamolecules by "2D" AFM nanomanipulation of highly uniform and smooth gold nanospheres

*Kyung Jin Park*[1,+], *Ji-Hyeok Huh*[1,+], *Dae-Woong Jung*[2], *Jin-Sung Park*[3], *Gwan H. Choi*[2], *Gaehang Lee*[4], *Pil J. Yoo*[2], *Hong-Gyu Park*[3], *Gi-Ra Yi*[2], and *Seungwoo Lee*[1,2*]

[1] SKKU Advanced Institute of Nanotechnology (SAINT), Sungkyunkwan University (SKKU), Suwon 16419, Republic of Korea, [2] School of Chemical Engineering, Sungkyunkwan University (SKKU), Suwon 16419, Republic of Korea, [3] Department of Physics, Korea University, Seoul 02841, Republic of Korea, [4] Korea Basic Science Institute (KBSI) and University of Science and Technology, Daejeon 34113, Republic of Korea

[+]Equally contributed to this work.
*Email: seungwoo.skku.edu

**Abstract**: Atomic force microscopy (AFM) nanomanipulation has been viewed as a deterministic method for the assembly of plasmonic metamolecules because it enables unprecedented engineering of clusters with exquisite control over particle number and geometry. Nevertheless, the dimensionality of plasmonic metamolecules via AFM nanomanipulation is limited to 2D, so as to restrict the design space of available artificial electromagnetisms. Here, we show that "2D" nanomanipulation of the AFM tip can be used to assemble "3D" plasmonic metamolecules in a versatile and deterministic way by dribbling highly spherical and smooth gold nanospheres (NSs) on a nanohole template rather than on a flat surface. Various 3D plasmonic clusters with controlled symmetry were successfully assembled with nanometer precision; the relevant 3D plasmonic modes (i.e., artificial magnetism and magnetic-based Fano resonance) were fully rationalized by both numerical calculation and dark-field spectroscopy. This templating strategy for advancing AFM nanomanipulation can be generalized to exploit the fundamental understanding of various electromagnetic 3D couplings and can serve as the basis for the design of metamolecules, metafluids, and metamaterials.

1. Introduction

Plasmonic metamolecules consisting of clustered metallic nanoparticles (NPs) (i.e., meta-atoms) have been extensively investigated over the last decade, because allowing for various electromagnetisms of interest at visible frequencies such as artificial magnetism, Fano-like interference, negative refractive index, and ultrahigh refractive index.[1–18] As accessible artificial electromagnetisms of plasmonic metamolecules are defined by materially realizable clusters, there is a pressing need for innovation of NP assembly methods. In this context, atomic force microscopy (AFM) nanomanipulation has been viewed as a unique and deterministic method for the assembly of clustered NPs, because its core, a blueprint containing information for programmed linear vector motion of the AFM tip can be directly translated into the directional dribbling of NPs with nanometer precision. This AFM nanomanipulation results in the assembly of plasmonic clusters with exquisite control over NP size, cluster geometry, and NP spacing, which would be difficult to achieve with other methods.[12,18,19] Nevertheless, AFM nanomanipulation has been relatively underutilized method in plasmonic and metamaterial engineer's toolset. This is in part due to its low robustness of NP dribbling and its limited ability to 3D cluster.

In general, the conventional synthesis of relatively large (> 20 nm) gold nanospheres (AuNSs) has resulted in a polygonal surface and highly dispersed size rather than smooth and uniform NSs.[20] Thus, dribbling polygonal-shaped and polydispersed AuNSs via AFM nanomanipulation has been error prone and difficult to achieve nanometer-precision.[21,22] Even nanometer error within gaps between AuNSs is non-trivial, as it leads to drastic changes in artificial magnetism.[9,12,18] Therefore, AFM nanomanipulation-enabled assembly of plasmonic metamolecules needs to address the relatively low robustness of AuNS dribbling. This lack of robustness has been effectively mitigated by recent advances in the synthesis of highly uniform and spherical AuNPs.[23–25] It has been shown that the use of highly spherical and smooth AuNSs can increase the accuracy of AFM nanomanipulation;[18,19] however, limited dimensionality to the planar 2D motifs (e.g., 2D tetramer) has remained an obstacle for expanding design space for accessible plasmonic metamolecules via AFM nanomanipulation.

In this work, we aimed to advance AFM nanomanipulation by dribbling highly uniform and smooth AuNSs on nanohole templates (referred herein to as templated AFM nanomanipulation) rather than on a flat surface, so as to greatly expand available dimensionality of plasmonic metamolecules from 2D to 3D (Figure 1). More specifically, the programmed AFM nanomanipulation allowed us to push AuNSs to the top edge of the nanohole (Figure 1a) and then drop AuNSs into the desired position at the bottom of the nanohole (see Figure 1b). Thus, the 2D plasmonic nanocluster (e.g., trimer) was first assembled within the nanohole; then, the remaining AuNSs placed on the top surface of polymeric nanohole can be further pushed onto the already assembled clusters (Figure 1c). Thereby, 3D plasmonic metamolecules were assembled even from the 2D linear vector motions of the AFM tip. Also, through dark-field spectroscopy (Figure 1d) together with electromagnetic simulations, the relevant plasmonic modes were fully rationalized.

2. Results and Discussion

2.1 Templated AFM nanomanipulation for the assembly of 3D plasmonic metamolecules

First, we developed the poly(methyl methacrylate) (PMMA) nanohole template by rigiflex nanoimprint lithography.[26] The PMMA nanoholes were 200 nm in diameter and 80 nm in height, as shown in scanning electron microscopy (SEM) and dark-field optical microscopy (DFOM) images (Figure S1). Particularly, the height of the nanoholes was designed to be comparable to the diameter of AuNSs, which were assembled at the bottom of each nanohole, such that the top AuNS could be fluently pushed onto the already assembled clusters.

As we previously reported, the selective etching of Au octahedral vertices allowed us to obtain highly uniform, ultra-smooth, and single crystalline AuNSs. In this work, we used 77 ± 3.2 nm and 100 ± 3.2 nm AuNSs, as presented in Figures 2a-b (more information about AuNSs is included in Figure S2-4).[23] The uniformity of our AuNSs were quantitated by statistical analysis of the aspect ratio.[24] Toward this direction, we carried out the algorithmic analysis of at least 200 AuNSs using custom code (Figure S4).[24] As shown in Figure 2c, the aspect ratio of both 77 nm and 100 nm AuNSs was nearly unity (less than 1.06); indicating that our AuNSs were highly uniform and smooth. In contrast, the AuNSs synthesized by the conventional seed-growth method[20] showed a highly polydispersed aspect ratio ranging between 1 and 2.5 (Figure S5).

We then deposited individually separated 77 nm AuNSs on the top surface of the PMMA nanohole template. To avoid trapping AuNSs within the PMMA nanohole, we used a dry transfer printing method (see Figure S6).[27] Figure 2d shows representative dark-field scattering spectra of several AuNSs printed onto the PMMA nanohole template. The wavelength of scattering peaks was consistent and self-evident of highly uniform and smooth AuNS. However, scattering intensities were slightly varied across the AuNSs; indeed, DFOM imaging revealed the non-uniformity of scattering intensities (Figure 2e) because the rough surface of PMMA nanoholes together with the partial embedment of AuNSs within the PMMA, caused by mechanical printing, affected the light scattering.

The AuNSs were then pushed along the different directions using the linear vector nanomanipulation of the AFM tip, which is highlighted by the red arrows in Figures 2e-g and the black arrows in Figures 2h-i. The first step toward 3D assembly was to push three AuNSs to the edge of the PMMA nanohole and drop them into the bottom of the PMMA nanohole, forming a 2D trimer within the PMMA nanohole, as presented in Figures 2j-k. In the second step, the remaining AuNSs on the top surface of the PMMA nanohole template were pushed to the desired position on the already assembled trimer (see Figures 2l-m), resulting in the successful production of a 3D plasmonic metamolecules (i.e., tetrahedral cluster) (Figure 2n). The scattering color change from greenish to yellowish further evidenced the structural evolution of the 3D assembly in the order of monomer, 2D trimer, and tetrahedral cluster (Figures 2e-g).

The key advantage of AFM nanomanipulation is that AuNSs can be individually controlled and located to any desired position with nanometer precision. Thereby, arbitrarily controlled geometry of 3D plasmonic metamolecules together with a sub-5 nm gap can be achieved, as highlighted in Figures 3a-d. For instance, after the assembly of the bottom trimer within the PMMA nanohole (Figures 2j-k), a fourth AuNS can be flexibly positioned on the centroid of the bottom trimer (Figure 3a) or on one of the bottom AuNSs (Figure 3b). Additionally, straightforward extensions to a tetragonal pyramid (i.e., AuNS on the centroid of tetramer, shown in Figure 3c) and a pentagonal pyramid (i.e., AuNS on the centroid of pentamer, shown in Figure 3d) were achieved.

The ability to specifically place AuNSs with nanometer precision is unique to AFM nanomanipulation and would be difficult to attain with other methods. For example, the colloidal self-assembly is mainly governed by a packing consideration or thermodynamics; thereby, available cluster geometries are in principle limited to symmetric cluster or random aggregate.[6,15,17,28,29] However, in case, the symmetry of plasmonic metamolecules is needed to be broken especially for enhancing artificial magnetism or inducing Fano resonance.[12,13,18] For e-beam lithography, each AuNP can be arbitrarily positioned with high flexibility, so as to form metamolecules with controlled asymmetry.[30] Nevertheless, it is difficult to obtain a few-nanometer gap, which is essential for strong capacitive coupling between particles. Moreover, top-down fabrication inevitably depends on evaporated polycrystalline metals with the surface or edge roughness, such that plasmonic resonance could be dimmed.

## 2.2 Optical magnetism of 3D plasmonic metamolecules

We next verified the optical properties of the assembled 3D plasmonic metamolecules using dark-field spectroscopy and numerical simulation (see Methods). Our plasmonic metamolecules differ in that the PMMA nanohole was used as the substrate rather than a flat layer. Therefore, we investigated the effect of the PMMA nanohole on the optical properties of plasmonic metamolecules (i.e., trimer) by numerical simulation (Figure S7). The PMMA nanohole was found to have negligible influence on the optical properties of the plasmonic metamolecules.

Once the assembly of the 2D symmetric trimer within the PMMA nanohole was confirmed through both AFM imaging and dark-field spectroscopy (Figures S8-S9), we placed a fourth AuNS onto the centroid of the already assembled trimer. As indicated in the numerical prediction, the scattering spectra of 3D symmetric tetrahedral cluster are more independent on the polarizations (*s*- and *p*-pol) of incident light (Figures 4a-b) compared with that of the 2D trimer (Figures S8-S9). Figure 4c shows the relative contribution of the electric/magnetic dipole (ED/MD) and the electric quadrupole (EQ) resonances to the total scattering cross section (SCS) (for *s*-pol), and Figures 4d-f highlight the nature of these resonances. The optical characteristics of the symmetric tetrahedral cluster for *p*-pol illumination are included in Figure S10. The broad scattering at 600 nm to 800 nm was caused by ED resonance, mainly confined within the first and second AuNSs (Figure 4d). Also, the optical magnetism corresponding to the scattering spectral shoulder at 800 nm ("bright" mode) was driven by 3D circulating displacement current and electric field along the $1^{st}$–>$3^{rd}$–>$2^{nd}$–>$4^{th}$–>$1^{st}$ AuNSs (Figure 4e); thereby, MD resonance was generated. This red-shifted MD peak relative to that of ED was consistent with quasistatic analysis.[6] Figure 4f displays the spatial distribution of the induced magnetic near-field and dipole moment resulting from such 3D circulating displacement current. A stronger coupling of the incident *H* field can be achieved with 3D plasmonic metamolecules compared to the 2D counterpart (e.g., trimer); therefore, the optical magnetism of the symmetric tetrahedral cluster was stronger than that of the 2D trimer.

The hallmarks of the experimentally measured scattering spectra including ED and MD resonances matched well with the theoretical predictions (Figures 4g-i), confirming that the symmetric tetrahedral cluster was well assembled with high structural fidelity. The MD was relatively weak compared with the ED; as such, the scattering shoulder from MD was smeared (Figures 4b-c and 4h-i) into a broad ED scattering at 600 nm to 800 nm. To isolate the MD response, we used the cross-analyzer in the light pathway of dark-field spectroscopy,[6] allowing for the clear observation of the enhanced optical magnetism of the 3D symmetric tetrahedral cluster (Figure 4i).

2.3 Engineering of magnetism by symmetry-breaking

Slightly broken symmetries of plasmonic metamolecules, which can be, for example, driven by just a few nanometer-deviations from perfectly symmetric geometry, are non-trivial, in that they induce magnetic-based optical Fano resonance or amplify the artificial magnetism.[9,14,18,31] In what follows, we demonstrate that broken symmetries can be implemented into 3D plasmonic metamolecules with nanometer precision by templated AFM nanomanipulation. In the first example of asymmetric 3D plasmonic metamolecules (Figure 5a), the two gaps between the first and second and first and third AuNSs in tetrahedral cluster were slightly increased from 1.5 nm to 3.0 nm, whereas the other gaps remained unchanged (1.5 nm).

As presented in Figure 5b, the measured scattering spectrum of the asymmetric tetrahedral cluster (for *s*-pol) significantly differed from that of the symmetric counterpart. In particular, a pronounced dip in ED scattering was visible around at 760 nm. This spectral dip originated from magnetic-based optical Fano resonance (i.e., destructive interference between bright ED and dark MD modes, referred to as optical bianisotropy), as theoretically analyzed in Figures 5c-f. Due to a near-field interaction (i.e., Fano resonance), some portions of ED resonance were directly coupled to the circulating displacement current and electric field along the $1^{st}$->$3^{rd}$->$2^{nd}$->$4^{th}$->$1^{st}$ AuNS pathway (Figure 5c), resulting in a strongly induced MD moment (Figure 5d).

The total cumulative phase shift during the interference between bright ED ($\pi/2$) and dark MD ($\pi/2$) via ED -> MD -> ED was found to be $\pi$ (Figure 5e), providing additional evidence of Fano resonance and confirming the destructive interference between the ED and MD modes.[9,14,18,31] Furthermore, the resonant wavelength of the MD matched that of the scattering spectral dip (see Figure 5f). The theoretical analyses predicted a slightly blue-shifted Fano resonance (725 nm) compared with the experimental measurement (760 nm), possibly due to the AuNSs used in the experiment being slightly larger than 77 nm (e.g., 80 nm). Meanwhile, "bright" MD resonance can be induced at around 800 nm (blue line, Figure 5f); however, its weakness obscures the hallmark in a broad ED.

Also, 3D cluster symmetry can be substantially broken by putting the fourth AuNS onto the center of one of bottom AuNSs instead of onto the centroid of the trimer (Figure 3b). The full optical characterizations of this asymmetric 3D cluster are included in Figure S11.

Finally, we heterogeneously assembled 3D plasmonic metamolecules composed of non-equivalent AuNSs, so as to break 3D cluster symmetry (Figure 6a). First, the symmetric trimer was assembled at the bottom of the PMMA nanohole using three 77 nm AuNSs (Figure 6b). Subsequently, a 100 nm AuNS was transfer-printed onto the top surface of the PMMA nanohole template and pushed onto the centroid of bottom trimer (Figure 6c).

Figures 6d and 6e present the theoretically analyzed total SCS and relative contributions of ED, EQ, and MD to the total SCS, respectively. ED was mainly confined in the first and second AuNSs, as with the symmetric tetrahedral cluster, so as to result in a broad scattering at 570 nm to 800 nm. However, compared with the symmetric tetrahedral cluster, the broken symmetries driven by non-equivalent AuNSs further boosted the strength of the induced MD via 3D circulating electric field ($1^{st}$->$3^{rd}$->$2^{nd}$->$4^{th}$->$1^{st}$ AuNSs, as shown in Figure 6f), in that the scattering spectral shoulder caused by the MD moment was elusive, even in the total SCS at 710 nm. Figure 6g shows the spatial distribution of the induced MD, which was found to be much stronger than that of the symmetric tetrahedral cluster (Figure 4f). The experimental scattering showed good agreement with the theoretical prediction (Figures 6h-i). In particular, the filtered scattering spectrum through the cross-analyzer further elucidated the enhanced MD at 710 nm (Figure 6i), probing the high structural fidelity of the assembled asymmetric tetrahedral cluster consisting of non-equivalent AuNSs.

## 3. Conclusions

We proposed a simple yet versatile method for expanding the available library of AFM nanomanipulation-enabled plasmonic metamolecules from 2D to 3D motifs. In particular, templated dribbling of highly spherical and smooth AuNSs (termed as templated AFM nanomanipulation) was individually manipulated in a relatively deterministic way, allowing for AuNS assembly into 3D clusters with nearly arbitrary geometries. Nanometer-resolution of NP position and interparticle space, which would be difficult to attain with other conventional methods, was successfully achieved. As such, the peculiar electromagnetic behaviors at optical frequencies including artificial magnetism and magnetic-based Fano resonance were precisely tailored with the 3D plasmonic metamolecules. Furthermore, our method has potential to serve as an intriguing and generalizable tool for the control of other NP-based 3D couplings such as Rabi splitting and molecular optomechanics.[32,33]

## Methods

*Nanoimprinting fabrication of polymeric nanohole templates*. To obtain PMMA nanoholes with high structural fidelity, we employed rigiflex nanoimprint lithography using a high modulus elastomeric stamp based on polyurethane (PUA, 40 MPa modulus) rather than conventional polydimethylsiloxane (PDMS, 1.8 MPa modulus).[26] The silicon master pattern was obtained by a monolithic nanofabrication set comprised of e-beam lithography and reactive ion etching (RIE); the PUA mold was replicated from the Si master. The thin PMMA layer was coated onto the glass substrate and imprinted using the replicated PUA mold at elevated high temperature at 120 °C, which is higher than the glass transition temperature of the PMMA (115 °C).

*Dry transfer printing of AuNSs onto the top surface of the PMMA nanohole template*. The AuNSs were first dispersed onto a flat Si wafer using spin coating. AuNSs were individually separated on the Si wafer by adjusting the concentration of AuNSs solution and spinning rate. A flat PDMS stamp was then conformably contacted with the AuNSs-coated Si wafer using a high preload. The ratio of curing agent to precursor was 0.1. The rapid retraction of the PDMS stamp from the Si wafer completed the AuNS retrieval ("inking" process). The printing and subsequent slow retraction of the PDMS stamp allowed AuNSs to be printed only on the top surface of the PMMA nanohole template.

*Synthesis of highly uniform and smooth AuNSs*. We used the protocol described in our previous report.[23] We started with the synthesis of single-crystalline Au octahedra by a controlled reduction of chloroauric acid ($HAuCl_4$) in the presence of ethylene glycol, poly(diallyl dimethyl ammonium chloride) (polyDADMAC), and phosphoric acid ($H_3PO_4$). Then, the vertices and edges of the Au octahedra were selectively etched by adding chloroauric acid (i.e., oxidizing agent).

*AFM nanomanipulation*. A commercialized AFM system (NTEGRA spectra, NT-MDT) was used; the standard nanomanipulation protocol, provided by the NTEGRA spectra, was employed for programmed, linear vector motion of the AFM tip. To minimize the adhesion between the AFM tip and polyDADMAC (organic ligand of AuNSs), we coated platinum-iridium (Pt-Ir) on the surface of a silicon-AFM tip.[19] AFM nanomanipulation of the AuNSs was repeated until the desired dark-field scattering spectrum was obtained.

*Dark-field spectroscopy*. We used our custom-built dark-field spectroscope, in which optical microscope (Nikon Eclipse), imaging spectrometer (IsoPlane, Princeton Instruments), and a CCD camera (PIXIS-400B, Princeton Instruments) are integrated. The incident angle of the broadband light source was 64°; the polarizations were controlled to be *s*- or *p*-pol by a polarizer (Thorlabs Inc.). The scattered light was collected using an objective lens with a numerical aperture of 0.9 and was then analyzed with an imaging spectrometer and CCD.

*Numerical simulation*. Numerical simulation of plasmonic cluster SCS was carried out using the finite element method (FEM), which was supported by COMSOL Multiphysics software. A semi-infinite PMMA nanohole template was employed in the simulation model and a perfectly matched layer was designed to rationally encapsulate the substrate and cluster to minimize the reflection effect. ED, EQ, and MD in the multipole expansion were numerically computed (the required polarization vector was integrated inside the particles). Then, the integration of the complex Poynting vector (vector potential) allowed us to determine the relative contribution of ED, EQ, and MD to the total average of radiate power. Normalization of the obtained vector potential to the incident power resulted in SCS.[12] The Au dielectric constants were taken from Johnson and Christy.[34]

Acknowledgement

This work was supported by Samsung Research Funding Center for Samsung Electronics under Project Number SRFC-MA1402-09.

Supporting Information

Details for rigiflex nanoimprint lithography, highly spherical and smooth AuNSs, and optical analysis of plasmonic metamolecules are included in Supporting Information.

References

1. Alù, A.; Salandrino, A.; Engheta, N. Negative Effective Permeability and Left-Handed Materials at Optical Frequencies. *Opt. Express* **2006**, *14*, 1557−1567.

2. Brandl, D. W.; Mirin, N. A.; Nordlander, P. Plasmon Modes of Nanosphere Trimers and Quadrumers. *J. Phys. Chem. B* **2006**, *110*, 12302−12310.

3. Alù, A.; Engheta, N. Dynamical Theory of Artificial Optical Magnetism Produced by Rings of Plasmonic Nanoparticles. *Phys. Rev. B* **2008**, *78*, 085112.

4. Urzhumov, Y. A.; Shvets, G.; Fan, J.; Capasso, F.; Brandl, D.; Nordlander, P. Plasmonic Nanoclusters: A Path Towards Negative-Index Metafluids. *Opt. Express* **2007**, *15*, 14129−14145.

5. Alù, A.; Engheta, N. The Quest for Magnetic Plasmons at Optical Frequencies. *Opt. Express* **2009**, *17*, 5723−5730.

6. Fan, J. A.; Wu, C.; Bao, K.; Bao, J.; Bardhan, R.; Halas, N. J.; Manoharan, V. N.; Nordlander, P.; Shvets, G.; Capasso, F. Self-Assembled Plasmonic Nanoparticle Clusters. *Science* **2010**, *328*, 1135−1138.

7. Vallecchi, A.; Albani, M.; Capolino, F. Collective Electric and Magnetic Plasmonic Resonances in Spherical Nanoclusters. *Opt. Express* **2011**, *19*, 2754−2772.

8. Mühlig, S.; Rockstuhl, C.; Yannopapas, V.; Bürgi, T.; Shalkevich, N.; Lederer, F. Optical Properties of a Fabricated Self-Assembled Bottom-Up Bulk Metamaterials. *Opt. Express* **2011**, *19*, 9607−9616.

9. Sheikholeslami, S. N.; García-Etxarri, A.; Dionne, J. A. Controlling the Interplay of Electric and Magnetic Modes via Fano-like Plasmon Resonances. *Nano Lett.* **2011**, *11*, 3927−3934.

10. Mühlig, S.; Cunningham, A.; Scheeler, S.; Pacholski, C.; Bürgi, T.; Rockstuhl, C.; Lederer, F. Self-Assembled Plasmonic Core-Shell Clusters with an Isotropic Magnetic Dipole Response in the Visible Range. *ACS Nano* **2011**, *5*, 6586−6592.

11. Dintinger, J.; Mühlig, S.; Rockstuhl, C.; Scharf, T. A Bottom-Up Approach to Fabricate Optical Metamaterials by Self-Assembled Metallic Nanoparticles. *Opt. Mater. Express* **2012**, *2*, 269−278.

12. Shafiei, F.; Monticone, F.; Le, K. Q.; Liu, X.-X.; Hartsfield, T.; Alù, A.; Li, X. A Subwavelength Plasmonic Metamolecule Exhibiting Magnetic-Based Optical Fano Resonance. *Nat. Nanotechnol.* **2013**, *8*, 95−99.

13. Sheikholeslami, S. N.; Alaeian, H.; Koh, A. L.; Dionne, J. A. A Metafluid Exhibiting Strong Optical Magnetism. *Nano Lett.* **2013**, *13*, 4137−4141.

14. Yang S.; Ni, X.; Yin, X.; Kante, B.; Zhang, P.; Zhu, J.; Wang, Y.; Zhang, X. Feedback-Driven Self-Assembly of Symmetry-Breaking Optical Metamaterials in Solution. *Nat. Nanotechnol.* **2014**, *9*, 1002−1006.

15. Qian, Z.; Hastings, S. P.; Li, C.; Edward, B.; McGinn, C. K.; Engheta, N.; Fakhraai, Z.; Park, S.-J. Raspberry-like Metamolecules Exhibiting Strong Magnetic Resonances. *ACS Nano* **2015**, *9*, 1263−1270.

16. Lee, S. Colloidal Superlattices for Unnaturally High-Index Metamaterials at Broadband Optical Frequencies. *Opt. Express* **2015**, *23*, 28170−28181.

17. Höller, R. P. M.; Dulle, M.; Thomä, S.; Mayer, M.; Steiner, A. M.; Förster, S.; Fery, A.; Kuttner, C.; Chanana, M. Protein-Assisted Assembly of Modular 3D Plasmonic Raspberry-like Core/Satellite Nanoclusters: Correlation of Structure and Optical Properties. *ACS Nano* **2016**, *10*, 5740−5750.

18. Sun, L.; Ma, T.; Yang, S.-C.; Kim, D.-K.; Lee, G.; Shi, J.; Martinez, I.; Y. G.-R.; Shvets, G.; Li, X. Interplay Between Optical Bianisotropy and Magnetism in Plasmonic Metamolecules. *Nano Lett.* **2016**, *16*, 4322−4328.

19. Kim, M.; Lee, S.; Lee, J.; Kim, D. K.; Hwang, Y. J.; Lee, G.; Yi, G.-R.; Song, Y. J. Deterministic Assembly of Metamolecules by Atomic Force Microscope-Enabled Manipulation of Ultra-Smooth, Super-Spherical Gold Nanoparticles. *Opt. Express* **2015**, *23*, 12766−12776.

20. Perrault, S. D.; Chan, W. C. W. Synthesis and Surface Modification of Highly Monodispersed, Spherical Gold Nanoparticles of 50−200 nm. *J. Am. Chem. Soc.* **2009**, *131*, 17042−17043.

21. Kim, S.; Shafiei, F.; Ratchford, D.; Li, X. Controlled AFM Manipulation of Small Nanoparticles and Assembly of Hybrid Nanostructures. *Nanotechnology* **2011**, *22*, 115301.

22. Shi, J.; Monticone, F.; Elias, S.; Wu, Y.; Ratchford, D.; Li, X.; Alù, A. Modular Assembly of Optical Nanocircuits. *Nat. Commun.* **2014**, *5*, 3896.

23. Lee, Y.-J.; Schade, N. B.; Sun, L.; Fan, J. A.; Bae, D. R.; Mariscal, M. M.; Lee, G.; Capasso, F.; Sacanna, S.; Manoharan, V. N.; Yi, G.-R. Ultrasmooth, highly Spherical Monocrystalline Gold Particles for Precision Plasmonics. *ACS Nano* **2013**, *7*, 11064−11070.

24. O'Brien, M. N.; Jones, M. R.; Brown, K. A.; Mirkin, C. A. Universal Noble Metal Nanoparticle Seeds Realized Through Iterative Reductive Growth and Oxidative Dissolution Reactions. *J. Am. Chem. Soc.* **2014**, *136*, 7603−7606.

25. Kim, D.-K.; Hwang, Y. J.; Yoon, C.; Yoon, H.-O.; Chang, K. S.; Lee, G.; Lee, S.; Yi, G.-R. Experimental Approach to the Fundamental Limit of the Extinction Coefficients of Ultra-Smooth and Highly Spherical Gold Nanoparticles. *Phys. Chem. Chem. Phys.* **2015**, *17*, 20786−20794.

26. Choi, S.-J.; Yoo, P. J.; Baek, S. J.; Kim, T. W.; Lee, H. H. An Ultraviolet-Curable Mold for Sub-100-nm Lithography. *J. Am. Chem. Soc.* **2004**, *126*, 7744−7745.

27. Lee, S.; Kang, B.; H. Keum, Ahmed, N.; Rogers, J. A.; Ferreira, P. M.; Kim, S.; Min, B. Heterogeneously Assembled Metamaterials and Metadevices via 3D Modular Transfer Printing. *Sci. Rep.* **2016**, *6*, 27621.

28. Urban, A. S.; Shen, X.; Wang, Y.; Large, N.; Wang, H.; Knight, M. W.; Nordlander, P.; Chen, H.; Halas, N. J. Three-Dimensional Plasmonic Nanoclusters. *Nano Lett.* **2013**, *13*, 4399−4403.

29. Schade, N. B.; Holmes-Cerfon, M. C.; Chen, E. R.; Aronzon, D.; Collins, J. W.; Fan, J. A.; Capasso, F.; Manoharan, V. N. Tetrahedral Colloidal Clusters from Random Parking of Bidisperse Spheres. *Phys. Rev. Lett.* **2013**, *110*, 148303.

30. Ye, J.; Wen, F.; Sobhani, H.; Lassiter, J. B.; Dorpe, P. V.; Nordlander, P.; Halas, N. J. Plasmonic Nanoclusters: Near Field Properties of the Fano Resonance Interrogated with SERS. *Nano Lett.* **2012**, *12*, 1660−1667.

31. Fan, J. A.; Bao, K.; Wu, C.; Bao, J.; Bardhan, R.; Halas, N. J.; Manoharan, V. N.; Shvets, G.; Nordlander, P.; Capasso, F. Fano-like Interference in Self-Assembled Plasmonic Quadrumer Clusters. *Nano Lett.* **2016**, *10*, 4680−4685.

32. Benz, F.; Schmidt, M. K.; Dreismann, A.; Chikkaraddy, R.; Zhang, Y.; Demetriadou, A.; Carnegie, C.; Ohadi, H.; de Nijs, B.; Esteban, R.; Aizpurua, J.; Baumberg, J. J. Single-Molecule Optomechanics in "Picocavities". *Science* **2016**, *354*, 726−729.

33. Chikkaraddy, R.; de Nijs, B.; Benz, F.; Barrow, S. J.; Scherman, O. A.; Rosta, E.; Demetriadou, A.; Fox, P.; Hess, O.; Baumberg, J. J. Single-Molecule Strong Coupling at Room Temperature in Plasmonic Nanocavities. *Nature* **2016**, *535*, 127−130.

34. Johnson, P. B.; Christy, R. W. Optical Constants of Noble Metals. *Phys. Rev. B* **1972**, *6*, 4370.

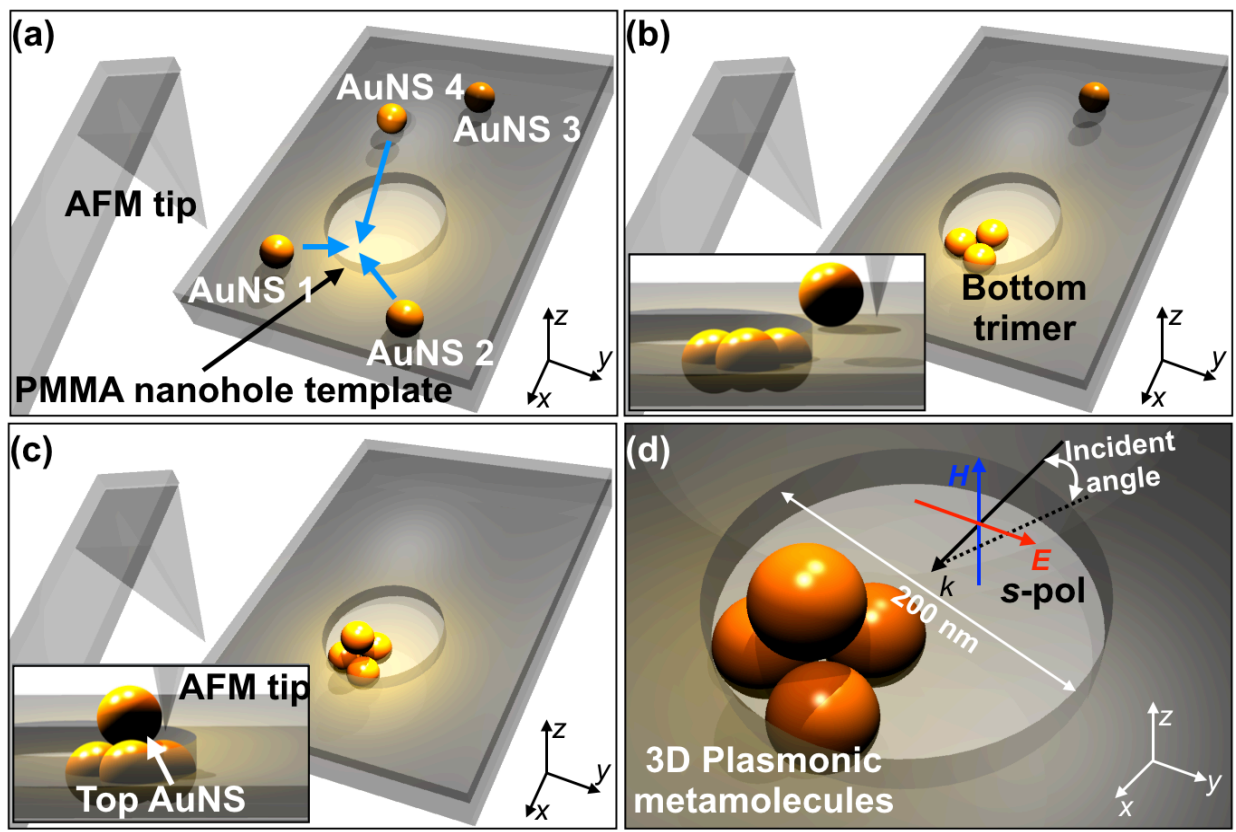

**Figure 1.** Schematic representation of the "3D" assembly of plasmonic metamolecules by "2D" vector nanomanipulation of an atomic force microscopy (AFM) tip. (a) Highly spherical and uniform gold nanospheres (AuNSs) were printed onto a poly(methylmethacrylate) (PMMA) nanohole template by dry transfer printing. (b) Three 77-nm-sized AuNSs were pushed and dropped onto the bottom of the PMMA nanohole via AFM nanomanipulation. (c) Remaining AuNS at the top PMMA nanohole template surface were pushed onto the already assembled trimer via AFM nanomanipulation. (d) 3D plasmonic metamolecules (e.g., tetrahedral cluster) were assembled and optically characterized.

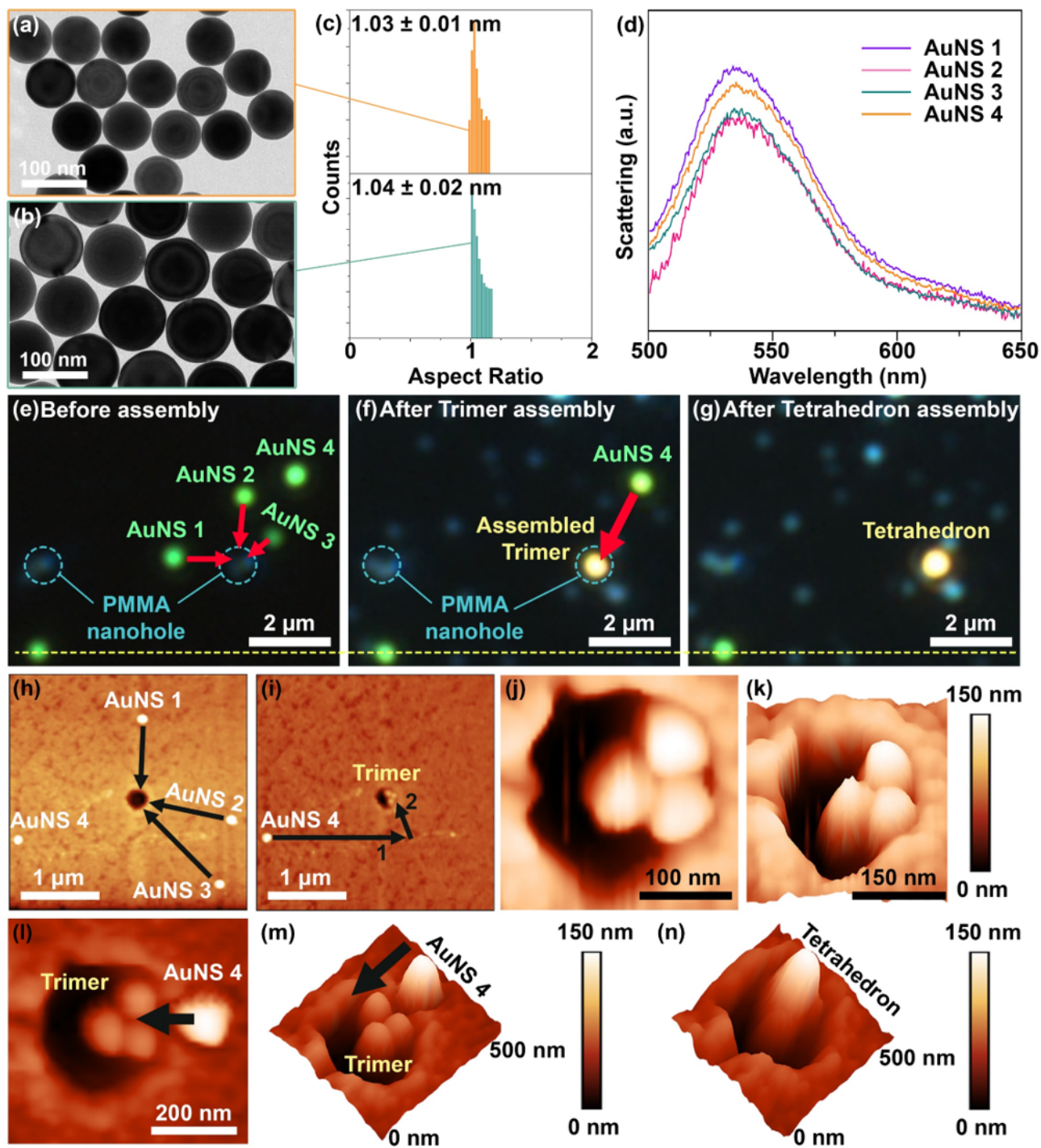

**Figure 2.** (a) Transmission electron microscopy (TEM) image of 77 nm AuNSs. (b) TEM image of 100 nm AuNSs. (c) Algorithmic analysis of size and shape uniformity of 77 nm (top panel) and 100 nm (bottom panel) AuNSs. Aspect ratio analyses allowed us to quantify the uniformity of AuNSs. (d) Dark-field scattering results of 77 nm AuNSs. (e-g) Dark-field optical microscope (DFOM) image of the stepwise assembly of 3D plasmonic metamolecules. (h) AFM image indicating transfer-printed 77 nm AuNSs and the AFM tip pathway used to assemble the trimer within the PMMA nanohole. (i) AFM image highlighting the AFM tip pathway used to push the fourth AuNS to the centroid of the already assembled trimer. (j-k) 2D and 3D AFM images of a trimer assembled within the PMMA nanohole. (l-m) 2D and 3D AFM images of 4$^{th}$ AuNS located at the edge of PMMA nanohole. The black arrow highlights the pathway of the fourth AuNS to complete the assembly of the tetrahedral cluster (representative 3D plasmonic metamolecule studied in this work). (n) 3D AFM image of the assembled symmetric tetrahedral cluster consisting of equivalent 77 nm AuNSs.

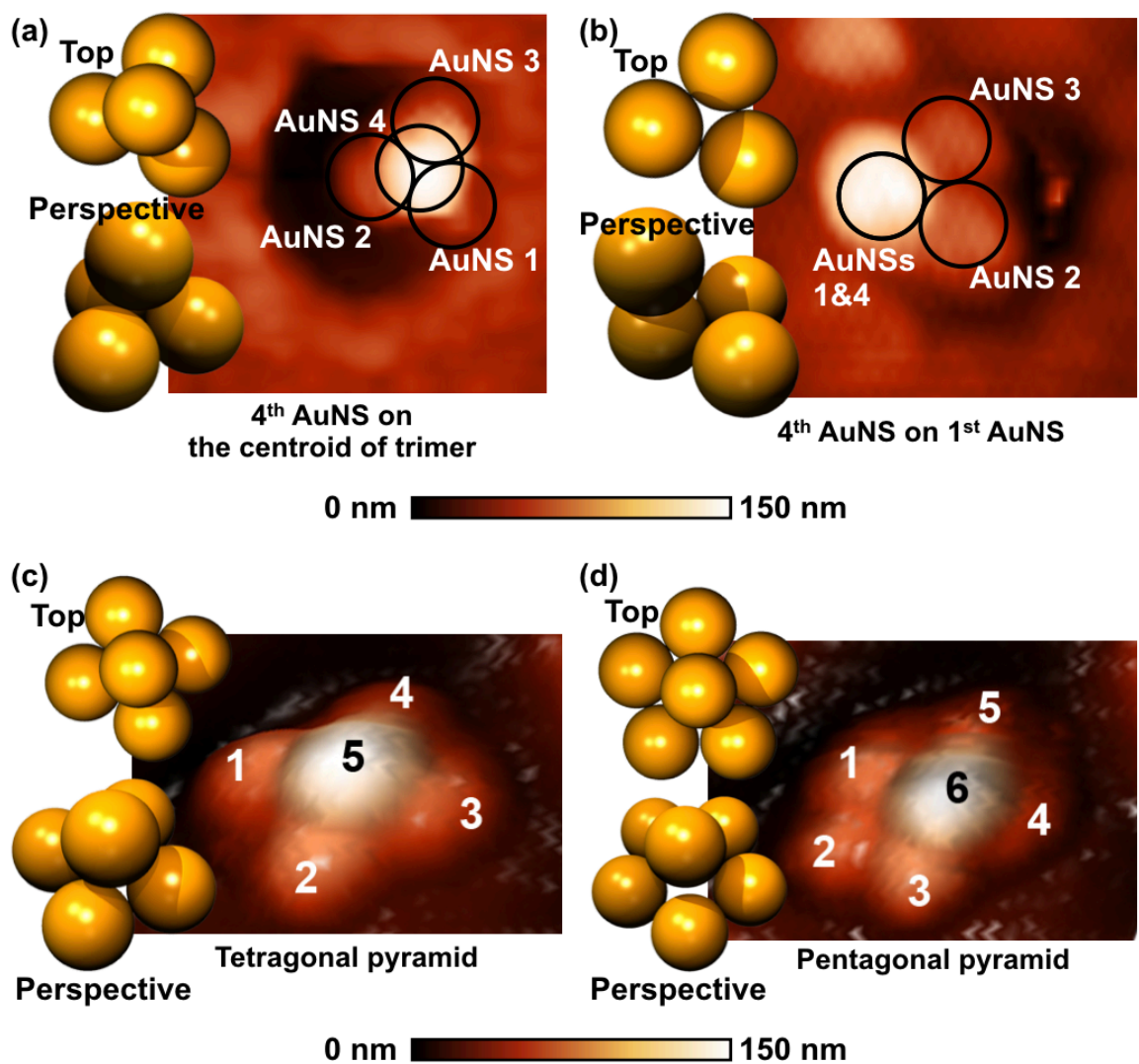

**Figure 3.** Representative examples of the 3D plasmonic metamolecules assembled via template AFM nanomanipulation. All these 3D plasmonic metamolecules are composed of 77 nm AuNSs. (a) 2D AFM image of the symmetric tetrahedral cluster. (b) 2D AFM image of asymmetric tetrahedral cluster (fourth AuNS located on the center of the first AuNS). (c) 3D AFM image of the 3D cluster with the tetragonal pyramid (i.e., 5th AuNS onto the centroid of the bottom tetramer). (d) 3D AFM image of the 3D cluster with the pentagonal pyramid (i.e., 6th AuNS onto the centroid of pentamer).

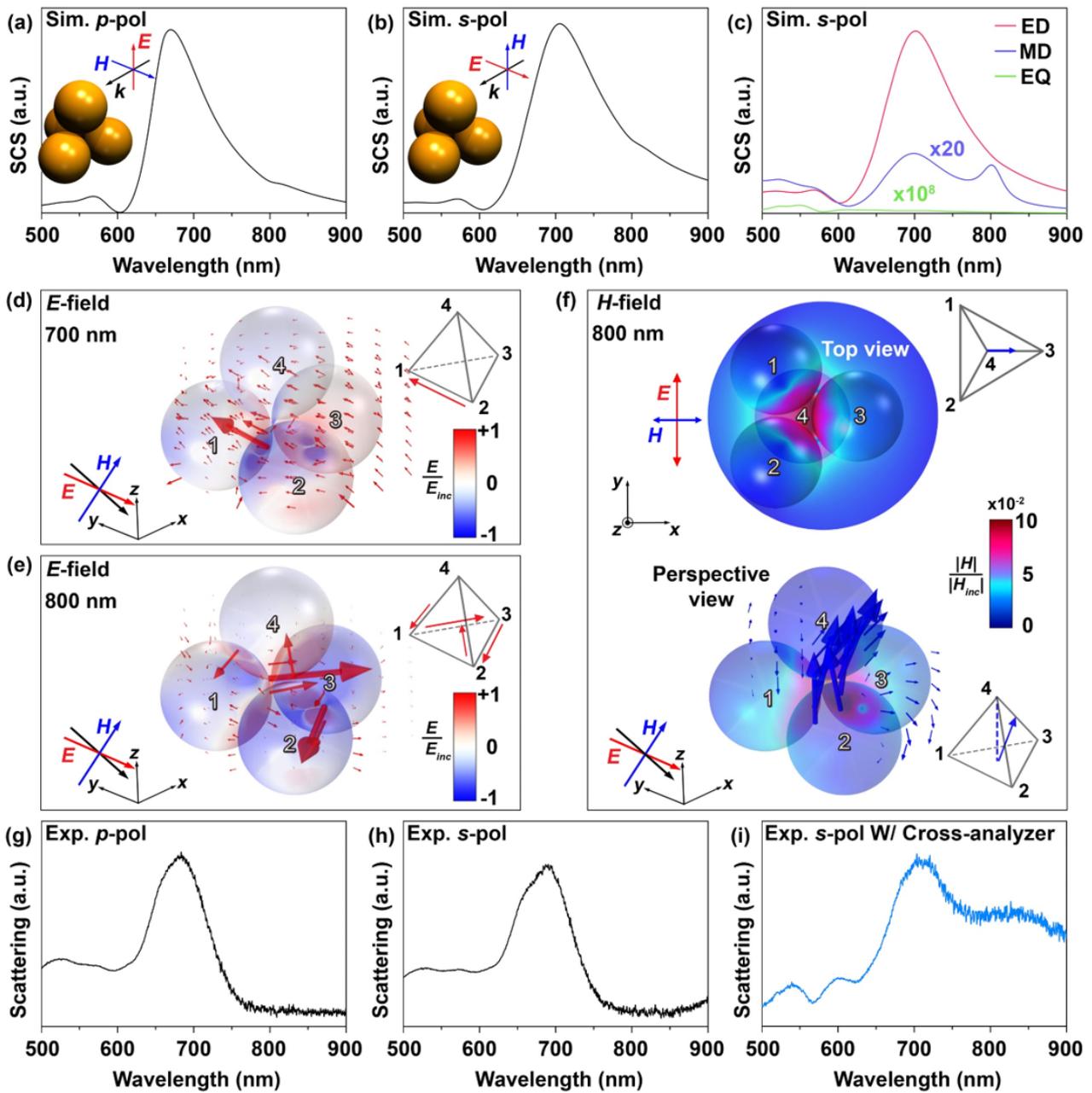

**Figure 4.** (a-b) The theoretical total scattering cross section (SCS) of the symmetric tetrahedral cluster for *p*- and *s*-pol illuminations. (c) The theoretical SCS analysis of the electric/magnetic dipole (ED/MD) and the electric quadrupole (EQ). (d) 3D spatial distribution of the electric field at 700 nm ED resonance. (e) 3D spatial distribution of the electric field at 800 nm MD resonance. The red arrows of tetrahedron frames, presented in the insets of (d) and (e), highlight the orientation of the induced electric field. (f) 2D (top panel) and 3D (bottom panel) spatial distribution of the magnetic field at 800 nm MD resonance. The blue arrows of tetrahedron frames (inset of (f)) indicate the main orientation of the induced MD. (g-h) The measured dark-field scattering for *p*- and *s*-pol illuminations. (i) *s*-pol dark-field scattering for the same tetrahedral cluster in (h), which was filtered by the cross-analyzer.

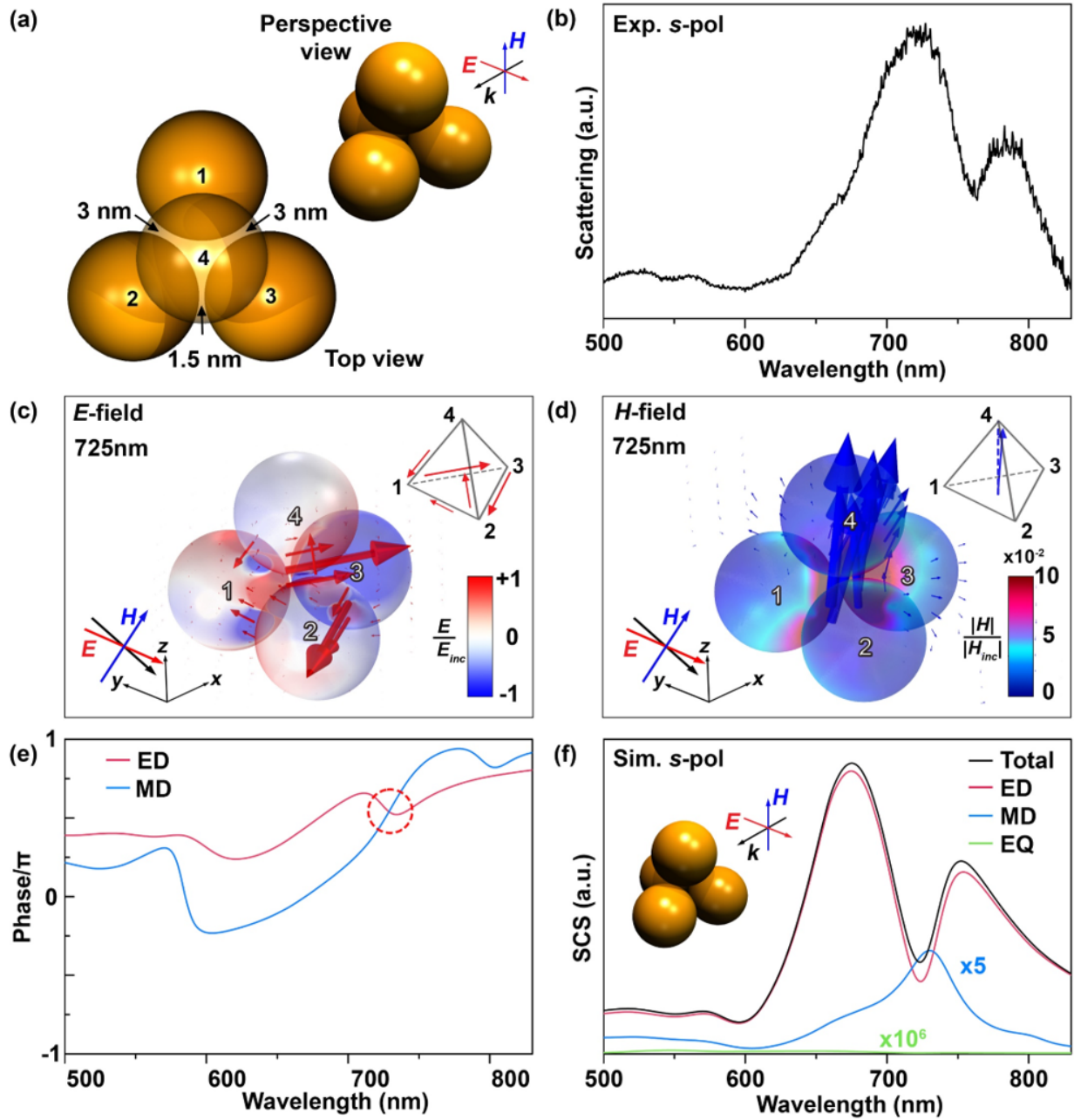

**Figure 5.** (a) Schematic representation of the asymmetric tetrahedral cluster composed entirely of 77 nm AuNSs. The broken symmetries of the tetrahedral cluster were implemented by small deviations of two gaps from symmetric geometrical frame (1.5 nm to 3.0 nm). (b) The measured dark-field scattering of the asymmetric tetrahedral cluster (a) for *s*-pol illumination. The magnetic-based optical Fano resonance was clearly evidenced by a distinct dip at 760 nm. (c) 3D spatial distribution of the electric field at the magnetic-based Fano resonance wavelength. The red arrows of tetrahedron frames, displayed in the inset of (c), indicate the orientation of the induced electric field. (d) 3D spatial distribution of the magnetic field at the magnetic-based optical Fano resonance wavelength. The blue arrows of tetrahedron frames in the inset of (d) indicate the main orientation of the induced MD. (e) The near-field phase of ED and MD. (f) The theoretical SCS of ED, MD, and EQ for *s*-pol.

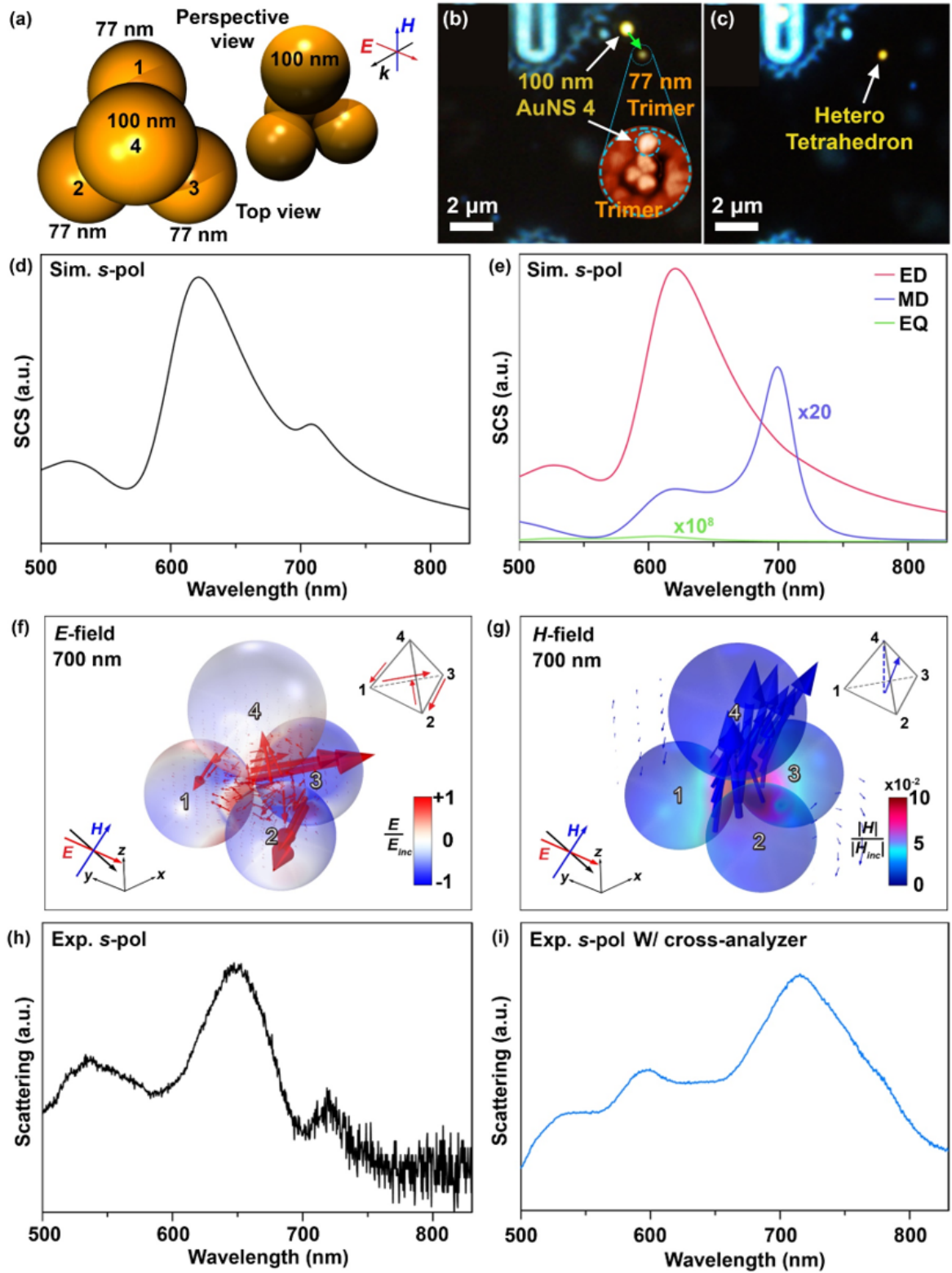

**Figure 6.** (a) Schematic representation of the asymmetric 3D tetrahedral cluster assembled with non-equivalent AuNSs (three 77 nm AuNSs and a single 100 nm AuNS). (b-c) DFOM of the stepwise assembly of asymmetric tetrahedral cluster in (a): (b) assembly of the bottom trimer with 77 nm AuNSs and (c) pushing the 100 nm AuNS onto the centroid of the already-assembled trimer. (d) Theoretical total SCS of the asymmetric tetrahedral cluster (a) for *s*-pol. (e) Theoretical analysis of the relative contribution of ED, QD, and MD to the total SCS. Spatial distributions of electric (f) and magnetic (g) fields at the resonance of optical magnetism (700 nm). (h-i) The measured dark-field scattering spectra of the asymmetric tetrahedral cluster (a) with (h) and without (i) cross-analyzer.

Table of Contents

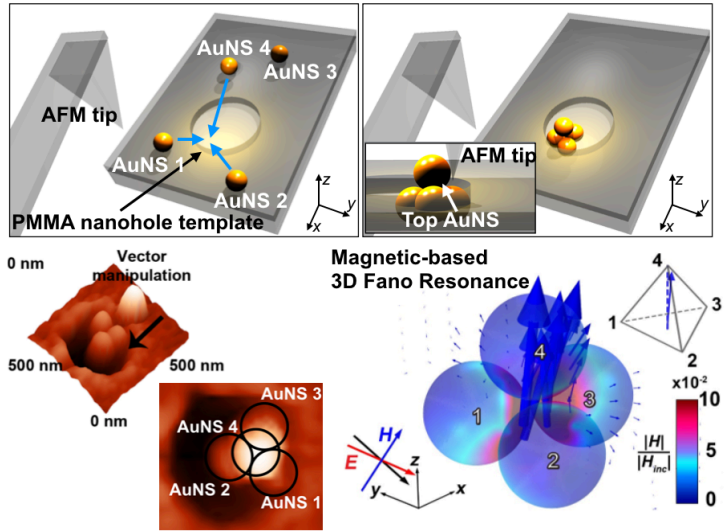